# Fundamental of cryogenics (for superconducting RF technology)

*Paolo Pierini*
INFN Sezione di Milano, Laboratorio Acceleratori e Superconduttività Applicata, Milano, Italy

**Abstract**
This review briefly illustrates a few fundamental concepts of cryogenic engineering, the technological practice that allows reaching and maintaining the low-temperature operating conditions of the superconducting devices needed in particle accelerators. To limit the scope of the task, and not to duplicate coverage of cryogenic engineering concepts particularly relevant to superconducting magnets that can be found in previous CAS editions, the overview presented in this course focuses on superconducting radio-frequency cavities.

## 1  Cryogenics and CAS

Several previous CERN Accelerator Schools (CASs) [1–3] have extensively covered many of the theoretical and technological aspects of cryogenics, which is the practice of reaching and maintaining low temperatures. Cryogenic engineering is the technical discipline needed to guarantee the operational environment of superconducting devices, such as the cavities and magnets for particle accelerators.

In this short contribution I will illustrate a few key concepts and provide a few "practical engineering" considerations concentrated on the case of superconducting radio-frequency (RF) linear accelerators. Superconductivity in magnets (and the necessary cryogenic implications) is another broad topic, somewhat less relevant to the topic of this course (high-power hadron machines) and extensively covered in previous CAS courses [1–3].

A separate contribution to this course (by H. Podlech) is dedicated to the systematic comparisons between the room-temperature normal conducting RF technology and the superconducting RF technology, while this contribution concentrates on cryogenic concepts in general and design considerations on cryostats and cryogenics for superconducting RF linear accelerators. This review is of course incomplete in many areas (such as the cryogenic instrumentation) and intentionally "light" on the most theoretical topics that could easily require an entire course.

## 2  Superconducting RF cavities and their cryogenic requirements

When a DC field is applied to a superconducting device, electrons condensed in Cooper pairs carry all of the current, and the electric resistance vanishes. All electrons are condensed into pairs only at $T = 0$ K, and the fraction of paired electrons decreases exponentially until the material reaches its critical temperature, above which a purely resistive behaviour is shown.

In the case of time-dependent RF fields and currents, dissipation takes place at all temperatures above 0 K, due to the fact that unpaired electrons feel the effect of the RF field and, differently from the frictionless motion of the Cooper pairs, generate currents leading to resistive losses.

In any superconducting radiofrequency (SCRF) resonator, which is a device aiming at creating a pattern of electric and magnetic fields in a region of space enclosed by a metallic boundary, power is dissipated on the metallic cavity walls, according to their surface resistance $R_s$ and to the surface magnetic field intensity:

$$P_{diss} = \frac{1}{2} R_s \int_S |H|^2 dS \qquad (1)$$

It is therefore from exploring the behaviour of the surface resistance with respect to temperature and frequency for the superconducting and normal conducting cases that the advantages of superconductivity can be assessed.

## 2.1 The advantages of RF superconductivity

RF fields at a frequency $f$ penetrate into a normal conducting metal (of electrical conductivity $\sigma$) by the skin depth $\delta$, thus leading to a surface resistance given by

$$R_s = \frac{1}{\delta \sigma} = \sqrt{\frac{\pi f \mu_0}{\sigma}} \qquad (2)$$

Dissipation in a normal conducting device therefore depends on material properties (through $\sigma$) and has a frequency dependence as $f^{\frac{1}{2}}$.

The Bardeen, Cooper and Schrieffer (BCS) theory of superconductivity can be used to develop expressions for the surface impedance of superconductors on the base of several fundamental material properties (as the mean free path, coherence length, etc.) [4]. For the case of Nb at temperatures below its critical temperature and frequencies below the terahertz range these complex expressions can be rewritten as the following "engineering" approximate formula:

$$R_{BCS}(T,f) [\Omega] = 2 \times 10^{-4} \frac{1}{T} \left(\frac{f}{1.5}\right)^2 \exp\left(-\frac{17.67}{T}\right) \qquad (3)$$

where $T$ is expressed in Kelvin and $f$ is expressed in gigahertz [4].

The ratio between Eqs. (3) and (2) is shown in Fig. 1, for frequencies up to 3 GHz and the two particular temperatures of 4.2 K (boiling temperature of liquid He at ambient pressure) and 2.0 K (subatmospheric liquid He in the superfluid phase). This plot shows that in the range between 300 MHz and 1.5 GHz the use of superconductivity can significantly reduce the power dissipated on the resonator walls by four to seven orders of magnitudes (depending on operating temperature and frequency).

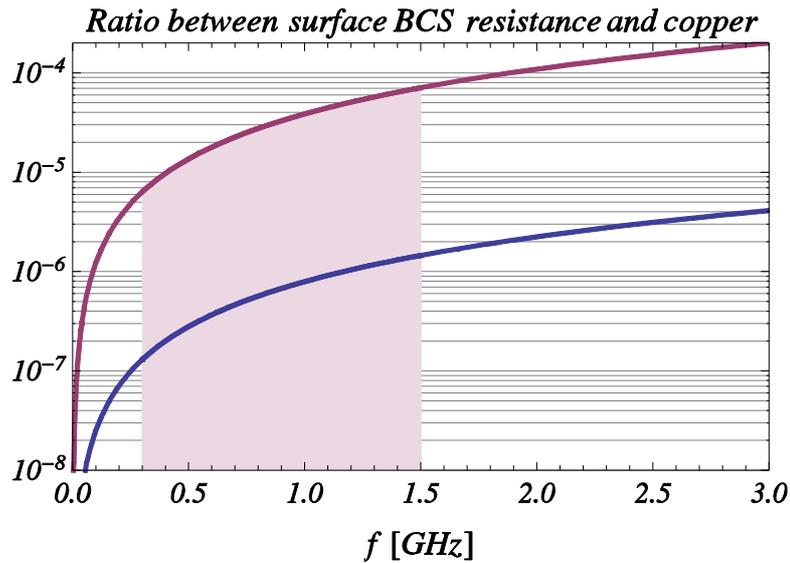

**Fig. 1:** Ratio between the BCS surface resistance term given by Eq. (3) to the copper surface resistance given by Eq. (2), for $T = 2$ K (bottom blue curve) and 4.2 K (upper purple curve). The "practical" SRF frequency region 300 MHz to 1.5 GHz has been highlighted.

## 2.2 SCRF Cavity fabrication technology

The situation illustrated by Fig. 1 represents an idealized condition, in real cases the surface resistance of a practically achievable superconductor is only partially due to the BCS contribution and can be written more generally as

$$R_s(T,f,H) = R_{BCS}(T,f) + R_{mag}(H,f) + R_{residual} \qquad (4)$$

where the first term after the BCS contribution represents the increased losses due to trapped DC magnetic fields (e.g. Earth's magnetic field, which needs to be properly shielded from the cavity environment) and the second is the so-called "*residual resistance*" contribution, accounting for several sources, such as chemical residues on the surfaces, foreign material inclusions, condensed gases or hydrides and oxides layers. To reach surface resistance values close to the BCS contribution, stringent material and procedures quality measures need to be implemented and strictly followed for the fabrication of a successful resonator made of a bulk superconductor material, such as Nb.

This is a broad topic, and here it is sufficient to mention that high-purity Nb material, free from inclusion and defects, should be used for the cavity RF surfaces. Clean joining technologies leading to no foreign material inclusions need to be followed during the resonator fabrication (most important is the electron beam welding technique). An aggressive chemistry, or electrochemistry, needs to be performed on all of the RF surfaces, to completely remove the surface layer damaged during the fabrication process, and all final treatments and cavity preparation for operation should take place in a clean room environment. A review of the SCRF Cavity technology can be found in Ref. [5].

## 2.3 The need for a cryoplant and the Carnot theorem

In addition to the cavity fabrication issues described above, one has also to realize that although superconductivity allows a dramatic decrease in the resonator power consumption, the power is deposited at the extremely low temperatures of operation.

Therefore, special measures are needed to achieve, guarantee and preserve the low-temperature environment required for the onset of superconductivity. In short, a *cryogenic infrastructure* (the cryoplant) is needed for the production and handling of the coolant, and for the removal of heat deposited at low temperatures.

The cryoplant is a thermal machine that performs work at room temperature to extract heat at low temperatures. Its two main functions are to bring the devices to the nominal temperatures and to keep them cool by removing any heat deposited at low temperatures. For a thermal machine that operates between the ambient temperature $T_{amb}$ and the cold temperature $T_{cold}$ to remove the heat ($Q_{in}$) from the cold temperature environment by means of performing work $W$ at the ambient temperature, the following relation holds:

$$Q_{in} \leq W \frac{T_{cold}}{T_{ambient}-T_{cold}} \qquad (5)$$

where the equality is valid for an ideal reversible process and the factor $T_{cold}/(T_{ambient} - T_{cold})$ represents the efficiency of the Carnot cycle between these temperatures. The inequality of Eq. (5) takes into account the efficiency of the real thermal machine, in which irreversible processes take place, which is in the range 25 % to 30 % for 4.2 K operation and 15 % to 20 % for 2 K. We can summarize these considerations in the following statements:

- to remove 1 W at 4.2 K, approximately 250 W are needed at room temperature;
- to remove 1 W at 2 K, approximately 750 W are needed at room temperature.

Even taking into account this overall thermal efficiency, however, Fig. 1 clearly shows that there is still an overall advantage for RF superconductivity in the moderate frequency region (up to ~ few gigahertz).

Niobium, with a critical temperature of 9.2 K, is the currently available material for fabrication of bulk superconducting resonators [4]. Table 1 shows the normal boiling point (temperature at which the liquid vapour pressure equals the atmospheric pressure) for various cryogenic fluids. Clearly, for the operation of SCRF resonators, the only available option is the use of helium.

**Table 1:** Normal boiling point of various fluids

|  | $_4$He | $H_2$ | Ne | $N_2$ | Ar | $O_2$ |
|---|---|---|---|---|---|---|
| Normal boiling point temperature, K | 4.22 | 20.28 | 27.09 | 77.36 | 87.28 | 90.19 |

### 2.4 Heat pumps as entropy pumps (get ready for non-idealities)

Entropy is the correct function of state that allows proper understanding and description of cryoplants, and the assessment of the non-ideality level of the process [6]. The Second Law of thermodynamics states that in any process the entropy output is always greater than the entropy input (and equal to that in the ideal, practically unreachable, reversible case).

Cryogenic systems can be described as entropy pumps [6], which transfer entropy from the cold region into the warmer environment (if we perform the necessary work). The top half of Fig. 2 represents a cryogenic plant as an entropy pump between the cold temperature, $T_{in}$, and the hot temperature, $T_{out}$, environments. To extract the heat flowing into the cold region, $Q_{in}$, entropy needs to be "transported" to the hot region and released there as a heat output at the hot temperature (as $Q_{out} = S_{out} \, T_{out}$).

On the top left part of Fig. 2 the ideal reversible process is illustrated. In this case entropy is conserved from the cold to the hot environments, thus the heat output is amplified by the ratio $T_{out}/T_{in}$. The energy conservation principle requires that the difference between the heat released into the hot environment and the heat deposited into the cold region is provided as work performed by the machine. On the top right part of the figure the irreversible case is shown, and the cryoplant is shown to introduce non-idealities, thus incrementing the entropy flowing from the cold to the hot environments. This increased entropy is then released as heat at the hot temperature level and additional work needs to be performed to make up for this non-ideality of real processes.

The bottom part of Fig. 2 shows a numerical example of the concepts expressed before for a cryoplant operating between 2 K and 300 K. A heat load of 1 W is deposited at 2 K, corresponding to an input entropy of 0.5 W/K. In the ideal case this entropy is preserved to the output and a heat load of 0.5 W/K × 300 K = 150 W is released to the ambient. A work of 149 W is thus needed by the ideal reversible process. If we now assume that non-idealities in the real process are such that entropy is incremented by five times (i.e. we have an internal entropy source of 2 W/K), then the output entropy of 2.5 W/K at the highest temperature corresponds to a heat rejection in the environment of 750 W, nearly all to be provided as work by the machine.

The overall efficiencies referred in the previous paragraphs are therefore related to the fact that real large helium cryoplants have internal entropy sources accounting for a multiplication of the incoming entropy from the load by a factor of ~3 at 4.2 K, and by a factor of ~5 at 2 K.

In a real cryoplant there are a large number of entropy sources contributing to the reduction of efficiency with respect to the ideal reversible case and accounting for the factors discussed above. A detailed review is presented in Ref. [6], together with an analysis of the various thermodynamic cycles used in cryoplants and the description of their main technological components. The largest single source of entropy production is typically associated with the gas compression stages (e.g. to account for gas recooling in multistage compression schemes by discharging heat to the environment). Entropy can also be introduced in the expansion stages (by spurious heat loads by conduction or radiation or by warm gas leaks), in heat exchangers and by heat leaks into the cold parts of the plant (e.g. by instrumentation cables or gas leaks).

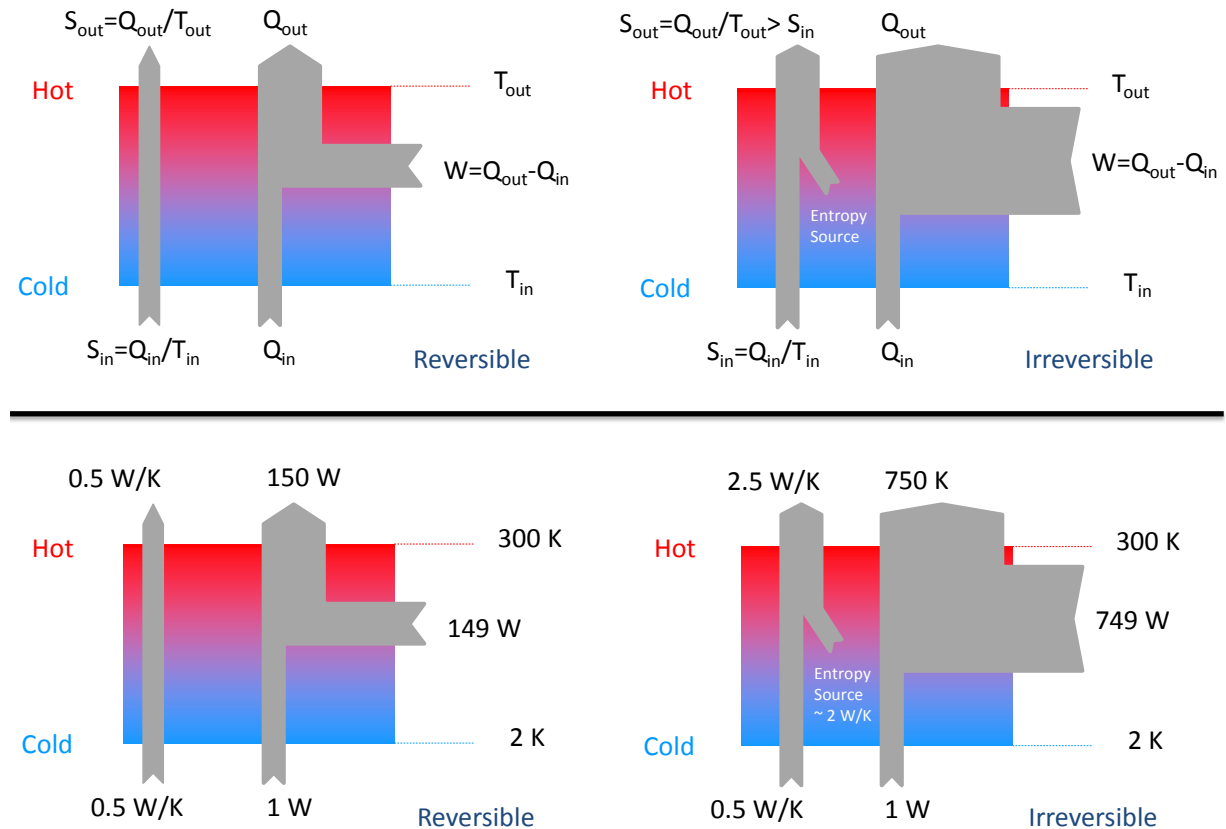

**Fig. 2:** Cryoplants as entropy pumps. The top part of the figure shows the entropy and heat flow for the ideal and irreversible cases. The bottom part illustrates a numerical example. For a full explanation of the figure refer to Section 2.4.

## 2.5 From a conceptual helium refrigerator to real machines

A conceptually simple cryogenic cycle aimed at removing a thermal load from a low-temperature region consists of two main stages: a *compression* stage, where the entropy content of the fluid is reduced at the expense of the work performed to compress the gas; and an *expansion* stage, which cools the fluid, either at the expense of internal forces or removing energy as work. In the compression stage the fluid is prepared to receive the entropy content of the load and entropy is then released to the environment by a heat exchanger right after compression. The expansion stage cools the fluid to a slightly lower temperature than the load, to extract entropy from it. These two stages are typically separated by a temperature staging device (as a counterflow heat exchanger) which guarantees that the two processes take place at different temperatures and prepares the fluid for optimal compression and expansion conditions. Heat exchangers allow transferring heat (i.e. entropy) from the cold region to the cycle and discarding it from the cycle to the environment.

The most common compound cycle used in modern cryoplants is the so-called Claude Cycle, displayed in Fig. 3 in its simplest variant, which can be optimized using several heat exchanges and expansion stages (typically using turbines) for efficiency.

A review of helium liquefaction plants and of the several variants of their cycle is outside the scope of this introductory course and the reader can find excellent coverage in specialized contributions of other CAS courses on superconductivity [6, 7], which also introduce all of the necessary thermodynamics concepts for their discussion.

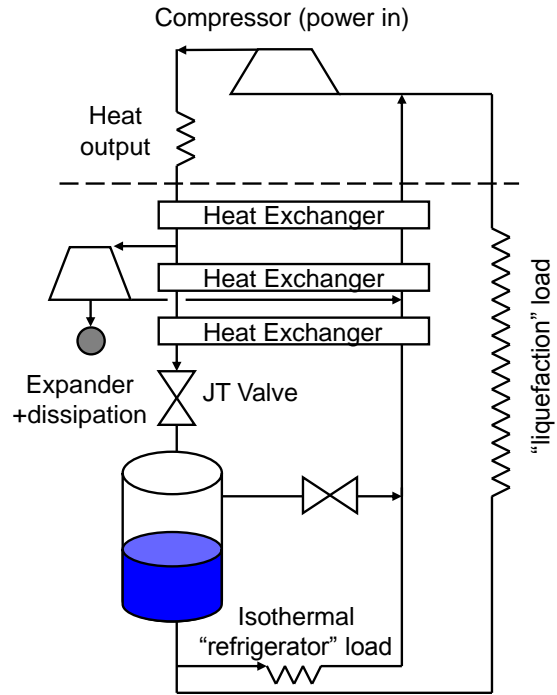

**Fig. 3:** Schematic view of a cryoplant based on the Claude cycle

## 2.6 Heat removal from the cold region: operating modes for a superconducting device

A major task of the cryoplant after cooling the superconducting devices is to guarantee the operating temperature by removing the heat deposited in the cold region. In general, heat is removed increasing the energy content of the cooling fluid (whether a liquid or a vapour). The cooling capacity is directly proportional to the fluid mass flow $\dot{m}$ and the enthalpy difference $\Delta H$ between the input and output fluid states:

$$P\ [\mathrm{W}] = \dot{m}\ [\mathrm{g/s}]\ \Delta H [\mathrm{J/g}] \tag{6}$$

Superconducting RF cavities are usually cooled in isothermal conditions either in pool boiling Helium I at atmospheric pressure (or slightly above, 4.2–4.5 K operation, as in HERA, LEP or KEKB) or in a saturated Helium II (superfluid) subatmospheric bath (2 K operation at 31.29 mbar, below the 2.17 lambda point, as in CEBAF, TTF, SNS, and the foreseen mode for ILC and ESS). In this isothermal cooling mode the heat absorbed by the load is spent in the phase transition from the liquid phase into the vapour phase (latent heat is 20.3 J/g for pool boiling He I at 1 atm, 4.2 K and 23.4 J/g for saturated Helium II bath at 31 mbar, 2 K).

The pressure–temperature phase diagram for He is shown in Fig. 4, where the typical regions of cooling mode of superconducting devices are highlighted. Figure 4 shows also cooling modes for superconducting magnets [9], usually cooled by forced flow of sub-cooled or supercritical He I (such as Tevatron, HERA and the SSC) or by pressurized He II (LHC). For magnets the operation with pressurized helium (single phase, either He I or superfluid He II) gives the maximum penetration of the coolant into the magnet coils, for increased heat transfer and stability, but leads to the need for a proper handling of the temperature rises induced along the cooling channels. For the thin-walled superconducting resonators, however, the pool boiling mode (especially in the case of a subatmospheric bath), in addition to the obvious advantage of a lower operating temperature [see

Eq. (3)], has the benefits of increased pressure stability, large local heat transfer capabilities to accommodate hot spots and isothermal cooling conditions. Saturated Helium II low-pressure (~30 mbar) operation is particularly important for the case of high-loaded quality factor (i.e. small bandwidth) resonators, to minimize field phase and amplitude perturbations induced by pressure variation in the bath.

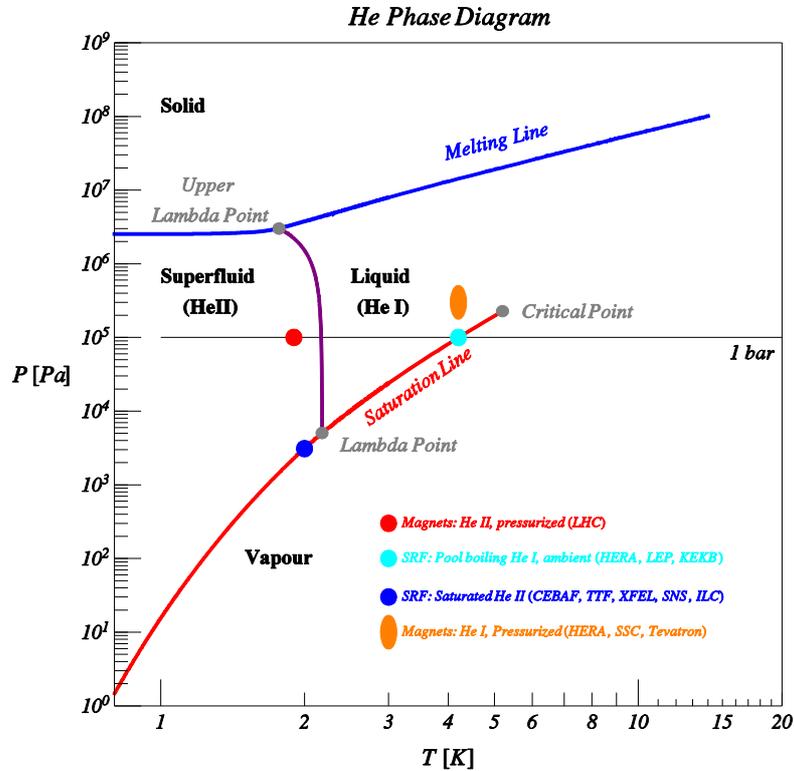

**Fig. 4:** Pressure–temperature phase diagram of helium

On the basis of the considerations expressed above, saturated He II operation with a two-phase mixture along the saturation curve is usually the preferred operation mode of high-field high-quality SCRF resonators. This operation mode, however, introduces a few additional complexities: a pumping system to establish the low-pressure operation; the need to handle pressure increase in the cryogenic piping at large mass flows not to perturb the operation temperature; the potential risk of air leaks into the low-pressure environment; and the operation of RF and electrical feedthroughs close to the minimum of the Paschen curve (e.g. in vertical test stations) where breakdown conditions are greatly enhanced.

Ultimately, heat is extracted from the cold region by evaporation of vapours from the saturated bath, which are then carried away. Since the latent heat of helium is quite small (~23.4 J/g at 2 K, for reference $N_2$ is ~200 J/g at 77 K), large heat deposition in the saturated He II bath imply large mass flows of low-pressure gas, i.e. a large volume flow. To reduce the volume flow large installations include cold compressors to increase pressure conditions of the cold gas before it reaches room temperature [8]. These devices also reduce the need of subatmospheric piping in the system, thus decreasing the possibility of air inleaks.

## 3      Cooling (and maintaining at cold) accelerator components

Physicists and engineers designing prototypical superconducting accelerator components usually concentrate mostly on the component design (either cavities or magnets) and later "jacket" them into helium vessels and cryostats for their testing.

For a large superconducting accelerator facility, however, the cooling mode of operation, the heat transfer mechanism in the operational environment, provisions for cooldown and warmup procedures and transient operation conditions need to be considered early in the component design and integrated into its supporting infrastructure, i.e. the cryogenic system. All of these considerations can affect the complexity (or, conversely, the simplicity) of the cooling system and are needed for a proper trade-off optimization between component and support system design.

### 3.1 Heat transfer mechanisms

One important consideration in the design of superconducting accelerator components operating at cold temperatures is the proper understanding and handling of heat transfer processes, especially those not strictly related to the loads associated with the proper device function (e.g. RF losses in a cavity will always occur at the cold operating temperature). In particular, spurious heat leaks due to mechanical supports, ancillary equipment as RF couplers and cavity tuning devices, or cabling for instrumentation and diagnostics, should be adequately minimized.

As explained in the previous sections, removal of power deposition at cold temperatures implies much larger power consumption at room temperature. Extending the considerations on thermal machine efficiencies outlined in Section 2.3 at intermediate temperature levels we have the following (approximate) efficiencies:

- 1 W deposited at 2 K requires approximately 750 W at room temperature;
- 1 W deposited at 4.2 K requires approximately 250 W at room temperature;
- 1 W deposited at 70 K requires approximately 12 W at room temperature.

These rough considerations suggest that it is crucial to intercept any spurious thermal flux from the room-temperature environment before it reaches the coldest region. This consideration applies to all heat transfer mechanisms that can take place in the operational environment by conduction, convection or radiation. For a detailed overview of heat transfer mechanisms, refer to Refs. [10, 11].

#### 3.1.1 Conduction

Heat is transported by conduction mechanisms inside solid or stagnant fluids, by processes occurring at the atomic scale. Conduction obeys Fourier's law, stating that the heat flow $\dot{Q}$ through a surface $S$ of a material with thermal conductivity $k$ in the presence of a temperature gradient $\nabla T$ is given by

$$\dot{Q} = -k(T) S \nabla T \tag{7}$$

In the simple single-dimensional case and for a material of length $L$ and cross section $S$ where the two end points are placed at the temperatures $T_{hot}$ and $T_{cold}$ can write the conduction equation in its integral form:

$$\dot{Q} = \frac{S}{L}\left(\int_{T_{ref}}^{T_{hot}} k(T)dT - \int_{T_{ref}}^{T_{cold}} k(T)dT\right) = \frac{S}{L}(K(T_{hot}) - K(T_{cold})) \tag{8}$$

where $K(T)$ is the thermal conductivity integral (evaluated from a reference temperature $T_{ref}$), usually found in literature or available from specialized software packages containing material properties at cryogenic temperatures [12].

Thermal conductivity (and its integral) varies greatly with temperature and with material. Proper choice of materials and thermal intercept strategies for the conduction paths to the cold environment is therefore necessary in the design of the SCRF components.

SCRF cavities need several penetrations from the room-temperature operation to provide structural support and frequency regulation, to provide the RF power for beam acceleration, to damp and extract spurious RF components at higher frequencies that may lead to detrimental effect on the beam characteristics, and cables for diagnostics and control. Direct connections from the room-temperature environment to the cold region should be avoided and, when needed, proper use of low thermal conduction materials should be made, providing thermalization at intermediate temperatures to intercept the heat flux under more favourable conditions.

Figure 5 shows thermal conductivity as a function of temperature for a few materials commonly used in cryogenic applications, and Figure 6 shows the thermal conductivity integrals for the same materials (with $T_{ref}=2$ K) as a function of the higher temperature $T$.

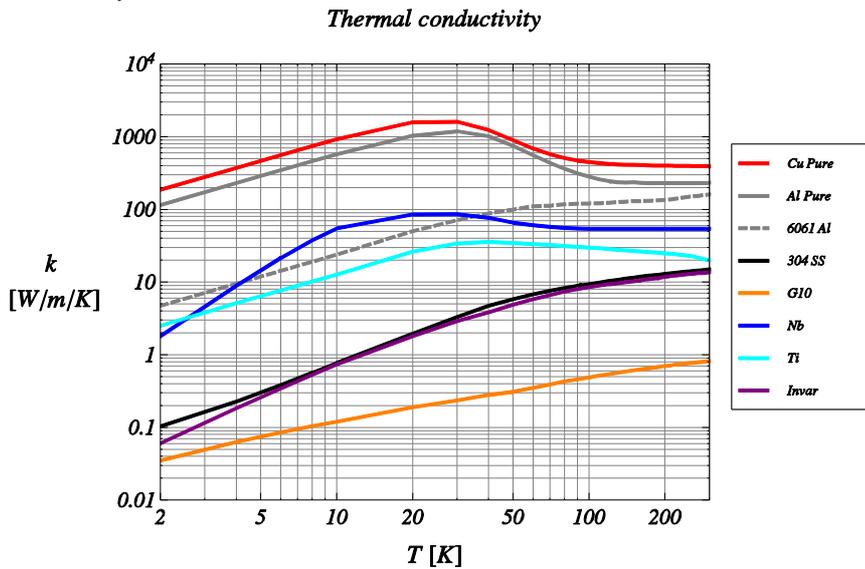

**Fig. 5:** Thermal conductivity of common materials used for cryogenic applications. Copper, aluminum and niobium curves refer to pure materials, showing the increased conductivity at the phonon peak. This characteristic is not shown by alloys (see e.g. 6061-T6 Al alloy). Data for this figure and Figs. 6 and 8 are from the CRYOCOMP package [12].

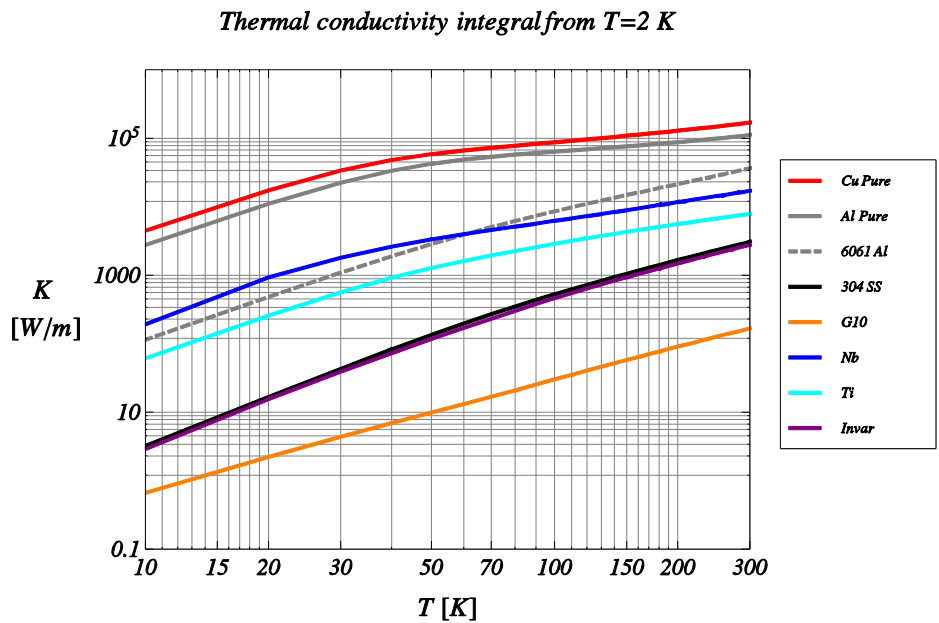

**Fig. 6:** Thermal conductivity integrals ($T_{ref}=2$ K) for the same materials as Fig. 5.

### *3.1.2 Convection*

Macroscopic fluid movement is the mechanism responsible of the heat transfer between the "wet" surfaces exposed to the fluid. The general law that can be used to describe convection heat transfer between a surface $S$ at temperature $T_{surface}$ with a bulk fluid at temperature $T_{fluid}$ is the Newton law of cooling [10, 11]:

$$\dot{Q} = hS(T_{surface} - T_{fluid}) \qquad (9)$$

where $h$ is the convection coefficient. In most situations it is impossible to derive an analytical formulation for the convection exchange coefficient, and the description of the heat transfer would require the numerical solution of partial differential equation associated with the fluid motion. Indeed Eq. (9) is a definition of the convection coefficient, which depends on the geometrical configuration of the flow and many fluid properties (its velocity, viscosity, thermal conductivity, specific heat and density). Determination of $h$ is usually performed from experimental empirical correlations between the dimensionless groups typically used in fluid mechanics to perform dimensional analysis. The most important of these dimensionless quantities are as follows (for extensive treatment, refer to Ref. [11].

The Reynolds number (*Re*), which represents the ratio between the inertia and viscous forces, and therefore allows the laminar ($Re < 2000$) and turbulent ($Re > 10^4$) regimes for flow conditions to be defined

$$Re = \frac{\rho v D}{\mu} \qquad (10)$$

The Nusselt number (*Nu*), which represents the ratio between the convection and conduction heat exchange mechanisms,

$$Nu = \frac{hD}{k} \qquad (11)$$

The Prandtl number (*Pr*), which is a property of the fluid, and represents the ratio between its ability to transport momentum and to transfer heat,

$$Pr = \frac{\mu C_p}{k} \qquad (12)$$

In these definitions $\mu, \rho, C_p, k, v$ indicate, respectively, the fluid dynamic viscosity, density, specific heat, thermal conductivity and fluid velocity evaluated at the fluid state, and $D$ is the characteristic dimension of the flow.

Empirical correlations, generally developed for non-cryogenic fluids, relate these parameters (in general expressed as $Nu = f(Pr, Re)$) for different geometry and flow configurations, and can be then used to determine the convection coefficient to be used in Eq. (9) to describe convective heat transfer. In general, except for the case of He II, the same correlations developed for non-cryogenic fluids can be used, with the caution to use them in their proper regions of validity (typically the flow regime and the geometrical configuration of the fluid/material interface), and to evaluate the fluid properties at the correct temperature and pressure conditions for the cryogen.

As an example, for turbulent liquid and gas single-phase internal flows the Dittus–Boelter correlation states that

$$Nu = 0.023 \, Re^{0.8} Pr^{1/3} \qquad (13)$$

So, to properly describe the convective thermal exchange in this case, first the turbulent flow regime needs to be assessed (by the calculation of $Re$ as given by Eq. (10)), then the fluid properties are used to evaluate $Pr$ and the value of $Nu$ is derived from the correlation expressed by Eq. (13). When $Nu$ is determined, the heat convection coefficient can be derived from the definition in Eq. (11) and the fluid properties. A number of correlations similar to Eq. (13) describe other typical geometrical and flow conditions usually found in cryogenic systems [10, 11].

Convection is one of the physical mechanisms by which we are able to extract heat from our devices and route it to the cooling fluids in the cryogenic piping. Analysis of convection exchanges are therefore important to make proper provision for the cooldown and warmup procedure and for the good behaviour of the thermal intercept circuits needed to prevent conduction load to the cold mass. This is also important for the piping that needs to extract the incoming radiation from the cold mass thermal shields (see below). An insufficient heat transfer coefficient in the presence of high heat loads would drive the pipe surface temperatures to higher values to sustain the transfer (by Eq. (9)), leading to increased temperatures for the heat interceptions of the shields.

### 3.1.3 *Radiation*

Heat is transported in the form of electromagnetic radiation emitted by surfaces, in the absence of any supporting fluid or medium. The total radiation flow emitted in all directions impinging on a surface $S$ by a body of emissivity $\varepsilon$ at the temperature $T_{hot}$ is given by

$$\dot{Q} = \varepsilon S \sigma_{SB} T_{hot}^4 \qquad (14)$$

where $\sigma_{SB} = 5.67 \times 10^{-8}$ W m$^{-2}$ K$^{-4}$ is the Stefan–Boltzmann constant.

Therefore, unshielded thermal radiation from the room-temperature environment reaching negligible temperatures would cause a load of approximately 500 W m$^{-2}$. A surface at the liquid nitrogen temperature would still cause 2 W m$^{-2}$ of radiation load. A suitable strategy to prevent these loads to reach the cold environment is needed.

The flow collected by a surface $S$ of emissivity $\varepsilon$ and temperature $T_{cold}$ coming from a black parallel surface at temperature $T_{hot}$ is

$$\dot{Q} = \varepsilon S \sigma_{SB}(T_{hot}^4 - T_{cold}^4) \qquad (15)$$

this general formula can then be extended for different geometries and to account for surfaces of different emissivities, and formulas are given in Refs. [10, 11]. For the simple case of two parallel plates of different emissivities, the quantity $\varepsilon$ in Eq. (15) can be substituted by the combined emissivity factor $\varepsilon_{hot}\varepsilon_{cold}/[\varepsilon_{hot} + (1 - \varepsilon_{hot})\varepsilon_{cold}]$. Material, temperature and surface finishing have a strong impact on surface emissivity, as can be seen from Table 2 [10, 11].

**Table 2:** Total emissivity for different temperature for selected material and surface conditions [10, 11]

| Material | Emissivity | | |
|---|---|---|---|
| | $T$ = 4.2 K | $T$ = 77 K | $T$ = 300 K |
| 304 Stainless steel, as fabricated | 0.12 | 0.34 | |
| 304 Stainless steel, mechanical polish | 0.074 | 0.12 | 0.16 |
| Aluminum, electropolished | 0.04 | 0.08 | 0.15 |
| Aluminum, mechanical polish | 0.06 | 0.10 | 0.20 |
| Aluminum, 7 μm oxide layer | | | 0.75 |
| Copper, as fabricated | 0.062 | 0.12 | |
| Copper, mechanical polish | 0.054 | 0.07 | 0.10 |
| Copper, with black paint (80 μm) | 0.892 | 0.91 | 0.935 |
| Aluminum coating on Mylar (both sides) | | 0.009 | 0.025 |

Radiation effects can therefore be somewhat mitigated by a proper use of material and surface finish condition. A more important measure for the management of radiative load in cryostats is to intercept the thermal flux impinging on the cold surfaces from the room temperature environment with one (or more) thermal screens actively cooled at intermediate temperatures from the two surfaces. Typically in all cryostats for superconducting magnets and cavities a shield make of a good thermal conducting material (typically Cu or Al) intercepts the ambient radiation flux at temperatures in the range from 40 K to 80 K.

A second very effective measure (often combined with the thermal shielding) to protect the surfaces from radiation load is to wrap them with many "floating" radiation-cooled reflective screens, interposed between the hot and cold surfaces. This is the concept of multilayer insulation (MLI) [8], in which several (typically 10 to 30) foils of reflective aluminium (or aluminized/double aluminized polyester films) are separated by a thermal insulating spacer material (as glass-fibre or polyester or paper foils) and "wrapped" around the cold surfaces to decrease the impinging thermal flux. The packing density of the layers affects performances, but even more important are the installation procedures. In particular, it is important to avoid "holes" leaving a direct line of sight to the cold environment and to prevent thermal short circuits between the layers, which have an impact on the concept of "floating" screens.

With MLI insulation [8, 10] the radiative load can be brought to the following levels:

- 0.5 $W/m^2$ to 1.5 $W/m^2$ from the room temperature environment to the intermediate shield temperatures (~40–80 K),
- 0.05 $W/m^2$ to 0.1 $W/m^2$ from the thermal shields to negligible temperatures.

To achieve these values proper care need to be taken around the needed penetrations in the temperature shields that need to accommodate pumping lines, support, current leads, RF feed and cabling, avoiding the exposure through holes of a direct line of sight from the "blackbody" ambient temperature environment.

## 3.2 Putting it all together: the cryostat/cryomodule environment

The cryogenic considerations expressed in Section 1, the heat transfer mechanism described in this section and the further engineering consideration exposed in the following pages play a large role in the design of one important item in a superconducting RF linac: the cavity cryomodule.

### 3.2.1 Functions of a cryostat/cryomodule

Cryomodules are the *modular building blocks* of all superconducting linear accelerators and need to fulfil the following main functions:

(i) provide mechanical support for the cavities (and possibly focusing elements);

(ii) meet alignment tolerances according to beam dynamics specifications;

(iii) create and maintain (efficiently) the cold environment for the cavity operation.

The linac cryomodules are also an important part of the cryogenic plant for any linac operation, since they are the regions where the major heat loads are located.

A conceptual "cartoon" view for a cryomodule of a superconducting RF linac is illustrated in Fig. 7. The cavities, placed in the central region, are supported to the external enclosure (the vacuum vessel), which is put under vacuum to inhibit gas-driven convective and conductive phenomena. Several penetrations connect the cold mass containing the cavities to the external environment (RF couplers, diagnostic and instrumentation cabling). The cold mass is wrapped by one or more layers of thermal shielding, with the primary role of intercepting the thermal radiation reaching the cold temperatures and the secondary role of convenient "manifolds" to provide thermal intercept to the conduction paths represented by the penetrations. Finally, a number of circuits provide the flow of the coolants needed for maintaining the cold mass and shield temperature levels, with the correct fluid conditions and flow rates to remove the estimated heat deposition (according to Eq. (6)).

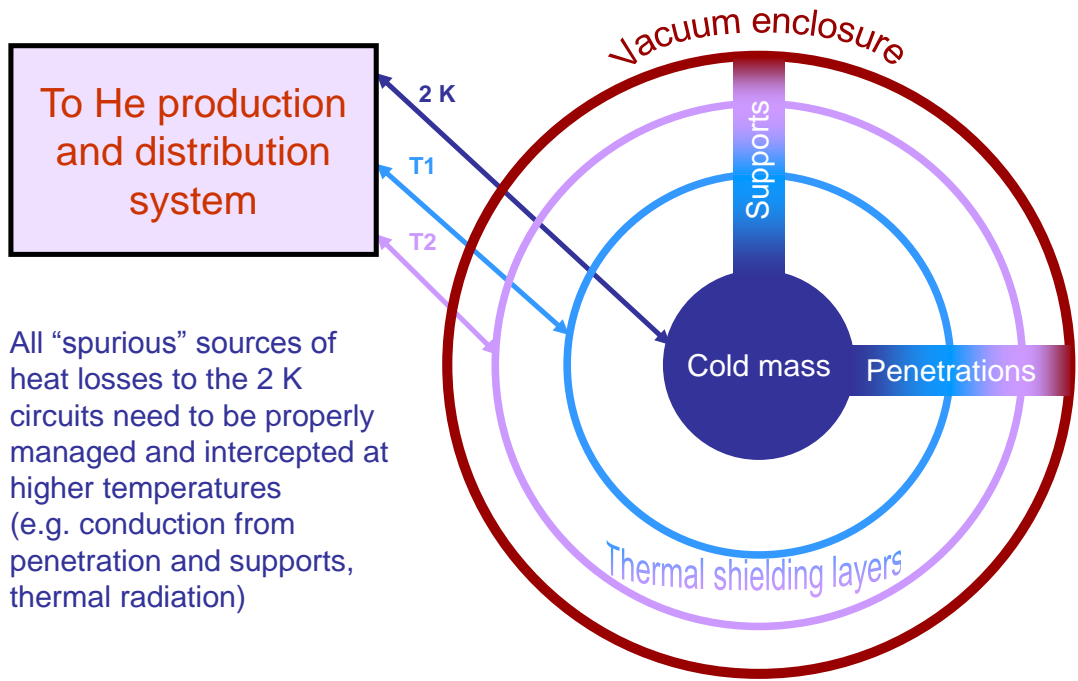

**Fig. 7:** A "cartoon" view of a SCRF cryomodule

### 3.2.2   *Engineering practices for cryomodule design*

Several engineering "practices" are therefore needed to develop a linac module concept, listed here along with their main goals:

- **Thermal design**
    - Minimization of spurious heat loads at the cold temperatures, especially important for large accelerator complexes.
    - Heat removal at various temperature levels, including provisions for thermal shielding and interception of the conduction paths.
    - Capabilities for cooldown and warmup, where the large enthalpy content of the cold mass needs to be carried away or restored in a short time without inducing large thermal gradient (and, thus, structural deformations).

- **Mechanical design**
    - Stable supporting of the cold mass, with minimal thermal losses.
    - Handling of gravity, vacuum and pressure loads.
    - Robustness with respect to thermal stresses induced by thermal gradients occurring during transient conditions and operation.
    - Provisions for implementation of reliable alignment of the sensitive components, and their preservation under differential thermal contractions.

- **"Hydraulics" and piping**
    - Integration of the cryostat cooling circuits in the cryogenic system design.

It is also important to note that the above tasks are not independent design actions that can be optimized independently from each other, since often the optimization strategies in the different domains would be conflicting. The most promising and stable mechanical support structure can very likely result in huge heat loads at the cold temperature, and therefore a coupled analysis and overall optimization is often required.

The following sections illustrate a few additional issues that need to be addressed in the design of cryogenic components.

### 3.2.3 Differential contractions

Designers of cryogenic components for accelerators face an additional complication given by the huge variation in the thermal contraction coefficients of different materials. Special care needs to be taken to account for this fact by selecting compatible materials when possible, allowing relative movements and avoiding to mechanical over-constrain the system, to prevent the occurrence of severe thermal stresses during cool-down, potentially capable of causing mechanical failures. Figure 8 shows the total linear contraction from room temperature as a function of the final temperature for selected materials. From this plot one can immediately see that at low temperatures:

(i) The only material with a behaviour similar to Nb is Ti (and this explains why the superconducting cavity helium reservoirs are often manufactured using Ti).

(ii) Stainless steel contracts twice as much as Nb, so relative displacement between the cold cavities and inner cryostat components is always a concern and needs to be handled.

(iii) Aluminum (pure and its alloys) has a thermal contraction 30 % higher than SS. Since the thermal shields are often made by Al or Cu for their good conduction properties, relative movement with respect to the cryostat structural components has to be allowed.

(iv) Special alloys (e.g. Invar) can be manufactured with very low thermal expansion coefficients. These material are important to provide "fixed" temperature-independent points for the positioning of critical components which need to have an interface at warm positions (e.g. in SCRF cavities the RF couplers connected to the vacuum vessel and bringing the RF power to the cold cavities).

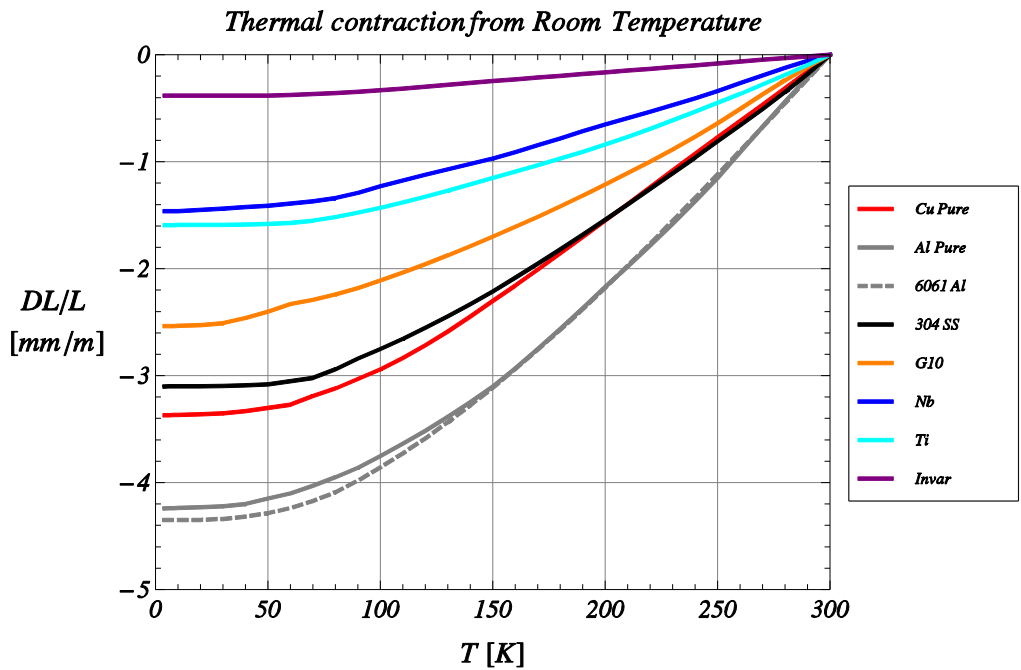

**Fig. 8:** Total thermal contraction from room temperature as a function of the final temperature level for different structural material for SCRF components

The effect of differential thermal contractions and relative component movement has to be taken fully into account in the design of the cryomodule, the container object for the superconducting cavities, as they have an impact on the final position of the components at the operating temperature, which needs to be compatible with the alignment accuracies required by beam dynamics analysis of the accelerator application.

### 3.2.4 Provisions for cooldown and warmup transients

When designing cryostats for superconducting devices, provisions have to be made for the necessary transient conditions to achieve operation: cooldown (and warmup). Very often the cryostat needs to provide separate helium lines for cooldown and filling operations. Filling lines from the top of the device can lead to interruptions of the liquid flow due to the outgoing vapours and either need to be placed at the bottom of the cold mass or require a separate line to vent the vapours.

Stratification of helium flow with a thermal gradient can lead to impractical long times for the cryostat warmup to ambient temperature, and electrical heaters or warmup lines to bring warm vapours need to be planned in the design.

### 3.2.5 Pressure drops

In a pipe of diameter $D$ and length $L$ where a fluid with uniform density $\rho$ and velocity $v$ flows at a mass flow rate $\dot{m}$ in the turbulent regime, neglecting elevation changes, pressure experiences a drop $\Delta P$ given by

$$\Delta P = \frac{8}{\pi^2} \frac{\dot{m}^2}{\rho D^5} L f \tag{16}$$

where $f$ is the fluid friction coefficient (depending on geometry, flow conditions and pipe roughness), which can be found tabulated in the literature [8].

Equation (16) is particularly important for the case of subatmospheric operation in He II, for the description of the pressure increase in the cryogenic circuit collecting the 2 K vapours from the cavities. The mass flow of the gas extracting the RF power deposited on the cavities can lead to a pressure increase at the cavity level, thus to a temperature increase (for operation along the saturation curve $T = T(P)$). An insufficient piping sizing would therefore lead to the inability to maintain the correct operating temperatures of the cavities, increasing dissipation according to Eqs. (1) and (3).

## 4 Case study: TTF/XFEL/ILC modules

One particularly significant example of a state-of-the-art cryostat for 2 K operation is the TTF cryomodule [13] (and its variants). Its design has been conceived in the 1990s for the concept of the TESLA superconducting linear collider and has been used for the construction of the Tesla Test Facility (TTF), now operating as the FLASH free electron laser user facility in DESY, Hamburg. The successful TTF design, with minimal modification, has been later adopted for the accelerator of the European XFEL Project, which will operate a ~2 km linac composed of 100 modules of this type. The concept has also evolved, with a few variations, into the baseline for the International Linear Collider (ILC) project, and adapted to other current or proposed projects (e.g. Cornell ERL, FNAL Project-X).

The module design is illustrated here to "summarize" the interplay of design issues at various system levels for accelerator/cryosystem and cryomodule.

### 4.1 The TESLA requirements

The TESLA 33 km collider proposal set the main initial requirements for the design of its modular block, as described in the following.

*4.1.1 High filling factor*

The energy reach of the TESLA 500 GeV to 800 GeV collider forced the maximization, to the maximum extent, of the ratio between the real estate gradient (i.e. total energy gain divided by overall accelerator length) and the cavity gradient performances (i.e. cavity energy gain divided by the nominal RF cavity length).

Thus, the design called for long cryomodules (containing many cavities) and to the proposal to connect them in long cryo-units, separated by short interconnections.

*4.1.2 Moderate cost per unit length*

Again, the scale of the project imposed a simple functional design, based on proven and reliable technology, readily available in the industrial context.

In particular, the cheapest allowable materials respecting operational load requirements have been selected, and the design effort has been directed to achieve the smallest number of machining steps per component.

To minimize operation costs, very small static losses are requested in the design (where static here means spurious losses to the cold environment in the absence of the dynamic loads due to RF cavity excitation). In particular, to avoid radiative load at 2 K levels, a double thermal shield has been foreseen in the design, with the outer shield operating around 70 K and an inner shield operating around 5 K. The double shields concept also allows a practical way to heat sink all conduction paths at these same intercept temperatures of the shield cooling circuits.

*4.1.3 Effective cold mass alignment strategy*

The module is designed to preserve, after cooldown, the alignment performed at room temperature. An extensive part of the engineering R&D phase at TTF has been dedicated to the assessment of the alignment reproducibility and stability, developing the necessary diagnostic methods and instrumentation, and performing the necessary experimental validation [14].

*4.1.4 Effective and reproducible assembly procedure*

The TESLA collider collaboration pushed the technology of bulk Nb superconducting cavities to unprecedented records. One important ingredient towards this achievement was the controlled cavity handling in a class 10 clean room up to the closure of the cavity string end gate valves. Thus, in the TTF assembly scheme the clean room preparation of the cavity string is completely separated from the module assembly. No cryomodule parts enter in the clean environment of the cavity assembly facility, thus allowing contamination of the cavity surfaces to be avoided or unnecessary expensive cleanliness requirements on cryomodule components to be imposed.

A reliable and cost-effective assembly scheme has therefore been implemented in the early stages of the development of the concept, leading to the definition of the necessary assembly toolings in parallel with the module design.

**4.2 Consequences and cryomodule concept**

The combined request of a high filling factor (to limit machine size) and low static heat losses (to limit operational costs) led to the integration of the cryomodule concept into the design and optimization of the overall cryogenic infrastructure of the collider. In particular, each cold–warm transition along the beam line and each cryogenic distribution box into the module require tunnel space and introduce additional heat losses. Thus, long cryomodules with many cavities (and focussing magnets) were preferred, cryogenically connected to form "cryo-strings" to minimize the necessity of cryogenic feeds. As a consequence, the helium distribution lines were integrated within the cryomodule environment. The TTF cryomodule contains eight cavities and a magnet package in ~12 m.

The limit of the length of each cryomodule unit is then set by its fabrication aspects (in terms of the need for large assembly tooling and precision machineries required for milling operations), module handling and transport considerations, and the foreseen capability to provide and guarantee the alignment within the required tolerances. Practically, this limit is in the 10 m to 15 m range.

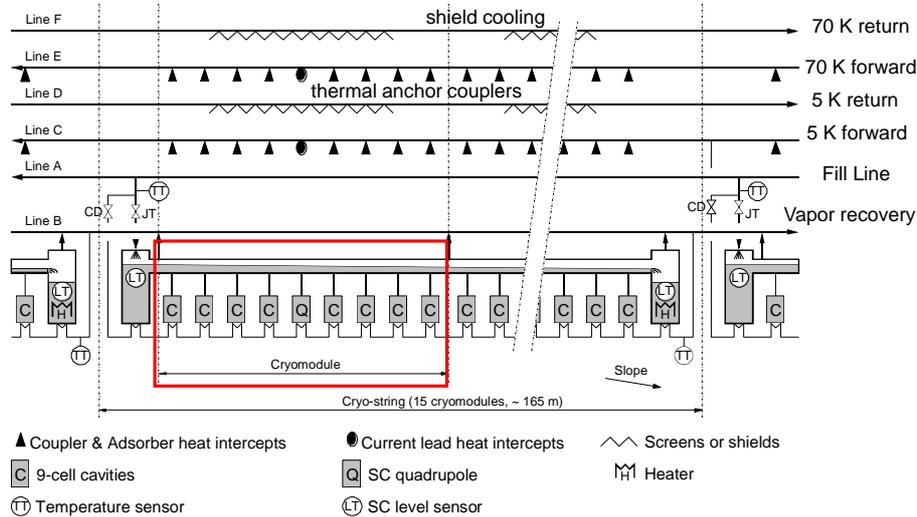

**Fig. 9:** Cryogenic circuits of the ILC cryo-string (courtesy of T. Peterson). The red box delimits the cavity and quadrupole components within one cryomodule.

The same concept of the TESLA [15] and XFEL cryo-strings briefly described above has been exploited for the cryogenic scheme of the ILC cryo-string [16], shown in Fig. 9. The various cryogenic lines that provide heat interception for the conducting paths (shown with a filled black triangle, on lines C and E) and thermal shields cooling (wavy lines, on lines D and F) are displayed in the upper part of the figure. The lower part of the figure shows that from the cryogenic point of view all cryomodules (one cryomodule unit is highlighted by the red box) are connected into a single continuous line of two-phase He II filling all of the cavities. This two-phase line is fed once per cryo-string by a subcooled pressurized He II line (line A). Vapour coming from the cavities is collected by a single gas recovery line (line B) running along the 12–15 module strings.

In this concept the cryogenic distribution for the whole cryo-string is integrated into the cryomodule, for static losses minimization. As the RF heat loads increase with the number of cavities in the modules and the cryogenic lines within the module serve the dozen or more modules connected in the cryo-string, the size of the cryogenic piping needs to be increased with respect to the case of a single, individually fed module (to guarantee the correct convective heat exchange and to contain the pressure drops). All of the cryogenic lines shown in Fig. 9 are integrated in the mechanical and thermal design of the cryomodule. Piping connections between adjacent cryomodules are welded, to guarantee continuity and avoid potential leaks from flanged connections.

To remove the RF power dissipated along one cryo-string formed by several cryomodules, a large mass flow of He gas is needed and, therefore, according to considerations expressed by Eq. (16), a large diameter is needed to reduce the pressure drop in the He gas return pipe (HeGRP). To combine a mechanical function, the HeGRP was dimensioned with an even larger diameter than strictly required by the handling of the pressure drop, so that it can act as the main structural backbone for the module string. Cavities and quadrupole are thus supported by this large structural backbone, which is stably thermalized at 2 K by the collected vapours.

The HeGRP is then supported by means of three low conductivity suspension composite posts to the vacuum vessel. The central support post is rigidly connected at the module vessel centre, whereas the two end ones are allowed to slide to recover differential contraction.

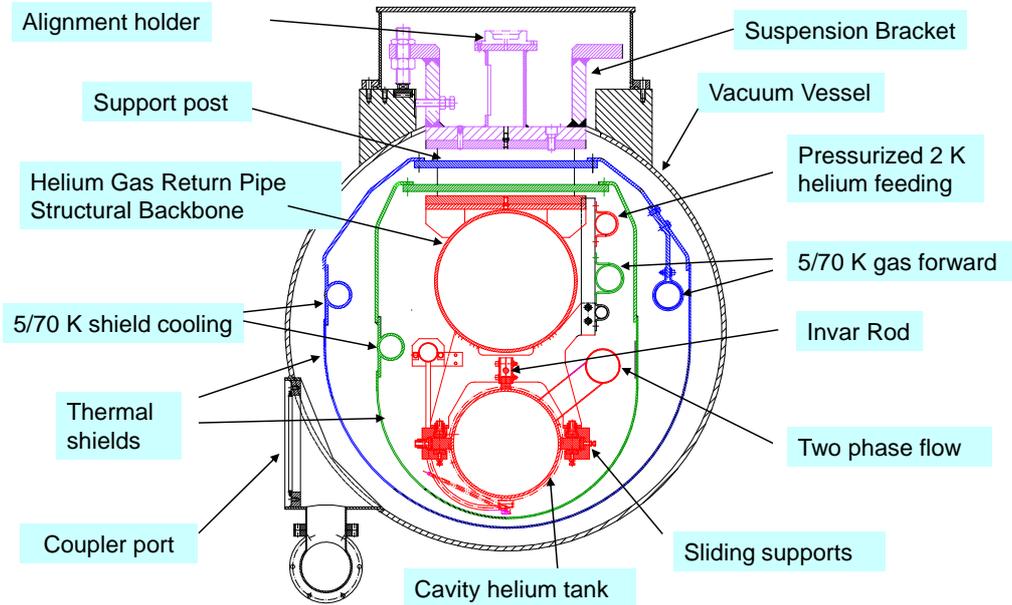

**Fig. 10:** The cross section of the TTF cryomodules

Figure 10 shows a cross section of the TTF/XFEL/ILC cryomodule concept, illustrating all of the characteristics described before:

(i) The large HeGRP supporting the string of eight cavities and a magnet package is clearly visible above the cavities in the 2 K region (shown in red). Below the HeGRP and above the cavity position the long invar rod provides longitudinal cavity fixing, while the cavity is supported by means of sliding supports [18]. To the right-hand side the two-phase pipe is used to fill the cavities with saturated He II at 2 K. At each module interconnection the He II vapours are brought to the HeGRP. The whole cold mass is supported by the composite G10 support post, and the conduction path is intercepted at the two thermal shield temperatures. The suspension brackets at the top of the support posts slide on linear rollers to accommodate differential contraction between ~12 m HeGRP and the vessel.

(ii) The two thermal shields at 70 K (blue) and 5 K (green), with integrated cooling pipes, protect the cold mass from the thermal radiation of the room-temperature vacuum vessel. Radiation load from the 5 K shield to the 2 K region is reduced to negligible levels. The thermal shields are cooled directly by an extruded aluminium pipe (shown in the left part of the picture), directly welded to the shield parts by means of a stress-relieving welding scheme [17]. The shields also act as thermal intercepts for the support posts and provide cable thermalization and effective coupler thermal intercepts via short thick copper braids.

(iii) The cryogenic lines bringing the cooling fluids forward through the modules downstream along the cryo-string (5 K and 70 K circuit and pressurized He II line) are shown in the top right part of the figure, and are thermally insulated from the 2 K cold mass by low thermal conductivity G10 supports. Heat load is absorbed by the pipe welded to the shields on the return path.

### 4.3 TTF/TESLA module achievements

In addition to the main design considerations summarized in this section, the design of the TTF/TESLA modules has provided engineered solutions for many of the items and cryostat design tasks outlined in the previous sections.

The cavity support from the HeGRP is realized by means of a sliding scheme [18] that allows the large differential contraction of the 12 m stainless steel HeGRP, completely decoupled from the cavities (which are built with Nb and Ti). The cavities, in fact, need to be fixed longitudinally at the warm position of the coupler ports on the vacuum vessel, to minimize stresses of the fragile ceramic RF window components in the couplers. Thus, a long invar rod (with a small coefficient of expansion) is used to clamp the cavity in this longitudinal position, which varies only slightly during the thermal cycle. Similarly, the support posts can slide on the vacuum vessel to allow the substantial length reduction of the HeGRP after cooldown.

The suspension and sliding cavity mechanisms, combined with a cold mass alignment strategy that relies on the cavity referencing to the HeGRP, has demonstrated the ability to achieve the necessary alignment reproducibility [14].

The thermal design has been verified by measurements of the integral heat loads of the module and fulfils the low static loads goals set by the TESLA Project.

## 5 Concluding remarks

This short lecture cannot cover all aspects of the cryogenic engineering concepts needed for the development of components for SCRF linacs, and a personal perspective has been offered here.

The main message for this lecture is that for a superconducting RF linac the design of its most critical component, the RF cavity, is only the starting point. A lot of physics considerations and detailed engineering need to be properly addressed to achieve the design of the supporting systems that need to provide its operating conditions. A key element of these systems is the cryomodule, the modular building block of the accelerator.

The overall design choices for the accelerator complex (especially for large machines) have strong constraints and implications on the cryomodule design, driving its conceptual definition, as reviewed in the case study.

Finally, plans for providing adequate mechanisms for cooling to nominal levels, heat removal during operation, control/preservation of alignment, countermeasures to prevent thermal stresses and avoid spurious heat leaks should be developed early in the cryostat and facility designs.

**Acknowledgements**

The author wants to thank all contributors to previous CAS courses dedicated to superconductivity in particle accelerators, in which the interested reader may find good coverage of topics only briefly mentioned here.

Particular thanks go to Carlo Pagani and many colleagues at INFN-Milano and at DESY-MKS, and to Tom Peterson, FNAL, for providing an endless repository of material to feed my curiosity and my slides.